\documentclass{article}
\usepackage[utf8]{inputenc}

\usepackage{graphicx}
\usepackage{algorithm}
\usepackage{authblk}
\usepackage{siunitx}
\usepackage{usebib}
\usepackage{xcolor}
\usepackage{ulem}

\usepackage{bm,bbm,amsmath,amsfonts,amssymb}
\newcommand{\indnt}[1][5]{\textbf{ }\hspace{#1mm}}

\title{Optimization of the injection beam line at the Cooler Synchrotron COSY
using Bayesian Optimization}
\author[1,2]{A. Awal}
\author[2]{J. Hetzel}
\author[2]{R. Gebel}
\author[2]{V. Kamerdzhiev}
\author[1,3]{J. Pretz}
\affil[1]{RWTH Aachen University}
\affil[2]{GSI Helmholtzzentrum für Schwerionenforschung}
\affil[3]{Forschungszentrum J\"ulich GmbH}

\begin{document}
\maketitle

\begin{abstract}
The complex non-linear processes in multi-dimensional parameter spa\-ces, 
that are typical for an accelerator, are a natural application for machine learning algorithms. This paper reports on the use of Bayesian optimization for the optimization of the Injection Beam Line (IBL) of the Cooler Synchrotron storage ring COSY at the Forschungszentrum Jülich, Germany. Bayesian optimization is a machine learning method that optimizes a continuous objective function using limited observations. The IBL is composed of 15 quadrupoles and 28 steerers. The goal is to increase the beam intensity inside the storage ring. The results showed the effectiveness of the Bayesian optimization in achieving better/faster results compared to manual optimization. 
\end{abstract}

\section{Introduction}
The field of accelerator physics has seen an increasing use of Artificial Intelligence (AI) and Machine Learning (ML) in recent years. An overview of using machine learning tools in accelerator physics is given in~\cite{edelen2018opportunities}. Bayesian optimization or algorithms have been used for example to optimize quadrupole settings for a Free Electron Laser~\cite{PhysRevLett.124.124801} for laser-plasma accelerators \cite{pousa} or tuning the emittance~\cite{Miskovich:2022erc}. In this paper we report on the optimization of a injection beam line connecting the cyclotron JULIC 
(JUelich Light Ion Cyclotron)
and the Cooler Synchrotron  COSY~\cite{cosy1,cosy2}.

The paper is organized as follows. The injection beam line is described in section~\ref{sec:inj}. The optimization algorithm is explained in section~\ref{sec:algo}. The results are discussed in section~\ref{sec:res}.

\section{The Injection Beam Line}\label{sec:inj}
\begin{figure}[h]
\centering
\includegraphics[width=.98\textwidth]{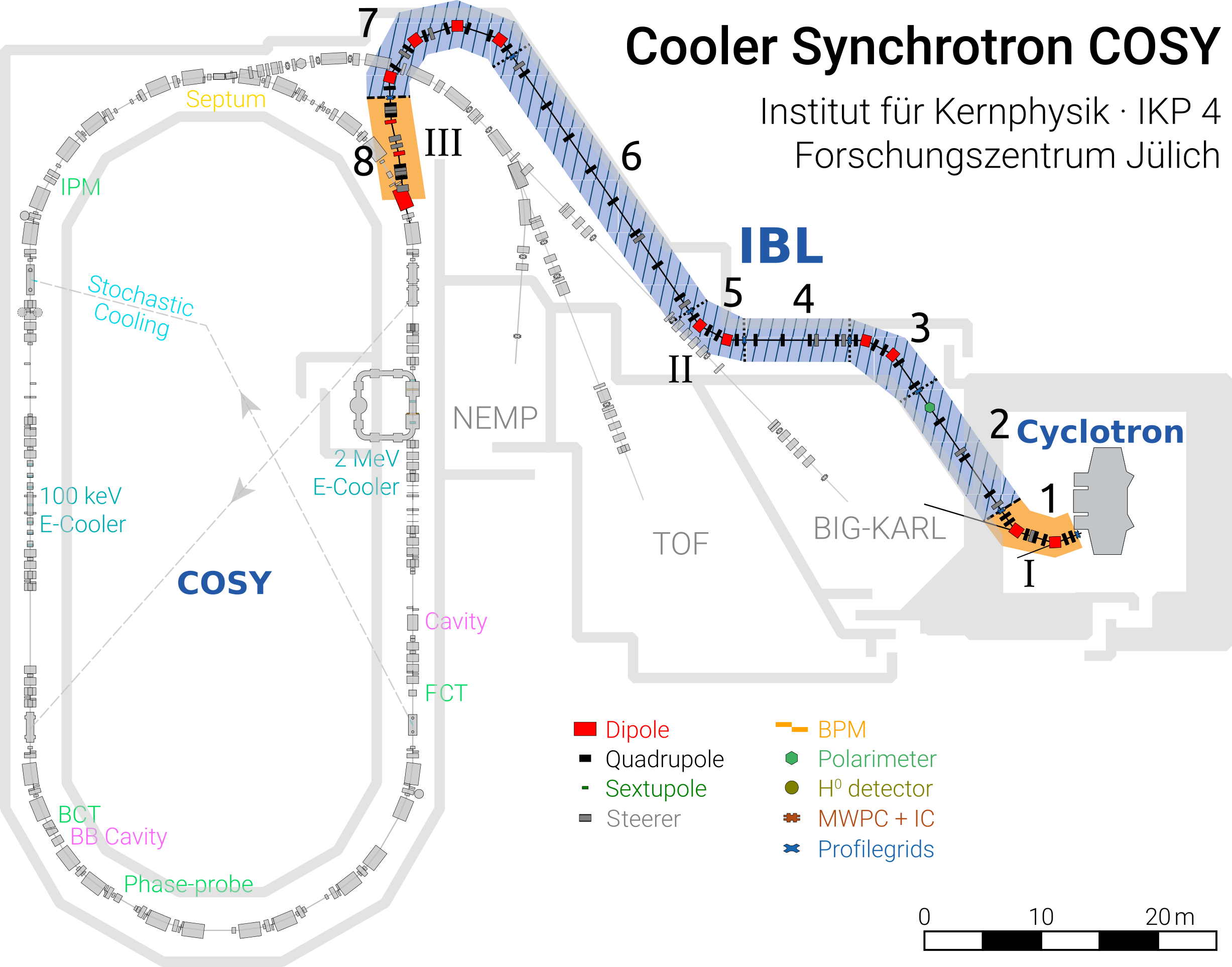}
\caption{The injection beam line (IBL) transfers polarized and unpolarized H-/D- ions from the cyclotron to the Cooler Synchrotron COSY.  At the end of the IBL the particles are injected into the storage ring via multi-turn stripping injection. The topical IBL is emphasised in the depicted overview of the COSY-facility. The division in sections is indicated by arabic numerals, whereas the segments are color coded and labeled with roman numerals.}%
\label{fig:cosy-ibl}
\end{figure}

The injection beam line (IBL) at COSY is the \SI{94.15}{\meter} long transfer line from the injection cyclotron (JULIC) to COSY (Figure~\ref{fig:cosy-ibl}). It transfers negatively charged hydrogen and deuteron ions with kinetic energies of \SI{45}{\mega\eV} and \SI{76}{\mega\eV}, respectively. For the experiment described here hydrogen ions were used. The electrons are stripped at the injection point in COSY via a stripping foil. In the storage ring, three fast bumper magnets move the orbit inside COSY horizontally to the stripping foil. Then the orbit bump is gradually reduced during the injection interval which lasts \SI{20}{\milli\second} within an injection cycle. The length of the injection cycle varies depending on the experiment. When optimizing the IBL, the time interval between two injections cycles was two seconds.

The IBL is composed of eight sections---four bent and four straight sections. These can be aggregated into three segments. The first segment receives the beam from the cyclotron and steers it into the second segment. Compared to the second segment it has more flexibility as it provides separate quadrupole magnets and steerers to transform and centralize the beam from the cyclotron. The second segment is the longest part of the IBL. It transports the beam from the first segment  to COSY and has less quadrupole families and steerers than the other two. The last segment provides the flexibility to transform the phase space of the beam to match with the acceptance of the storage ring at the injection stripping foil. There are three Faraday beam cups to measure the beam intensity; one at the beginning of the IBL and two between the three segments. These beam cups are usually deployed by operators to facilitate the optimization of the first two segments. There are in addition eight profile grids in both the horizontal and vertical direction. One is placed at the beginning of the IBL and the others are in between the eight sections of the IBL.
Table~\ref{tab:elements} gives an overview of all elements in the IBL.
\begin{table}[]
    \centering
    \begin{tabular}{|l|r|r|} \hline
        Component & nb. of elements & nb. of families \\ \hline
        Quadrupole & 43 & 15 \\
        Dipole & 12 & 5 \\
        Horizontal steerer & 15 & 
        - \\
        Vertical steerer & 13 & - \\ \hline
        Horizontal profile grid & 8 & - \\
        Vertical profile grid & 8 & - \\
        Faraday cup & 3 & - \\ \hline
    \end{tabular}
    \caption{Overview of the IBL elements. The upper section is for magnetic elements while the lower is for diagnostic instruments.}
    \label{tab:elements}
\end{table}

\section{Injection Optimization}\label{sec:algo}
As can be seen in table~\ref{tab:elements}, the IBL  has 15 quadrupole magnet families and 28 steerers summing up to 43 parameters to optimize. In this context a family is a group of magnets where all elements are operated with the same current and hence cannot be tuned independently. Dipoles have a fixed current and they are not part of the optimization procedure. Depending on the situation, different parts or magnet types of the IBL were optimized. The goal is to have the highest beam current inside the storage ring. 
\paragraph{Bayesian Optimization}\textit{ }\\
 Bayesian Optimization is a machine learning method that optimizes an objective function $f:\textbf{x} \in \mathbb{R}^d \rightarrow \mathbb{R}$ to find the optimum using limited observations. The objective function $f$ to be optimized is a black-box continuous function that is costly to evaluate and/or is difficult to differentiate. The number of dimensions $d$ for the input $\textbf{x}$ is typically not too large---typically $d\leq 20$~\cite{bo-tutorial}. As the dimension space of the IBL is 43, we decided to perform the optimization over sections or specific components. The goal is to find the maximum of the objective function with respect to the parameters vector~$\bf x$. In the scope of our problem, $\bf x$ is a vector of the magnet currents, and $f(\textbf{x})$ is the reading of the target measurement instrument such as the beam current inside the storage ring.
 
Bayesian optimization works by defining and computing two functions, a surrogate function and an acquisition function. The surrogate function approximates the objective function using the observations. It is computed first to model the probability distribution of the objective function in the unexplored space of parameters. Gaussian processes~\cite{gp} is a widely used surrogate function and it is the method used in our experiments. In Bayesian optimization, the surrogate function is then used to find the most promising parameters to sample based on an exploration-exploitation trade-off. Exploration promotes sampling from regions with high uncertainty while exploitation promotes sampling from regions with likely improvement over the current best observation. This is achieved through computing the acquisition function. The acquisition function adopted in the experiments is the expected improvement (EI)~\cite{ei}. EI is a widely used method that computes the expected value of improvement compared to the current best observation from the objective function. It is defined as

\begin{equation}
\begin{split}
\label{eq:ei}
EI(\textbf{x})=&\left\{\begin{array}{ll}
(\mu(\textbf{x}) - \mu^+ - \varepsilon) \varphi(Z) + \sigma(\textbf{x}) \phi(Z) & \text{,if } \sigma(\textbf{x})>0 \\ 
& \\
0 & \text{,if } \sigma(\textbf{x})=0
\end{array}\right. \\
&Z = \frac{\mu(\textbf{x})-\mu^+ - \varepsilon}{\sigma(\textbf{x})}\\
\end{split}
\end{equation}

\begin{figure}
\centering
\includegraphics[width=.8\textwidth]{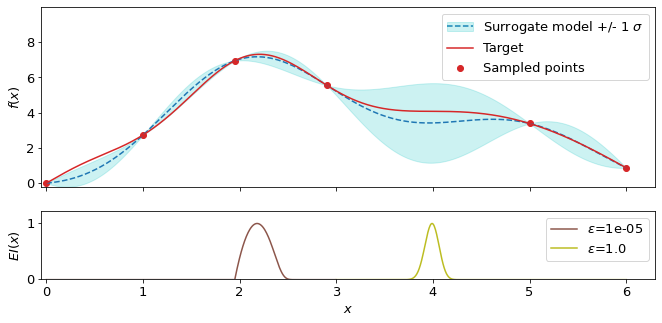}
\caption{An example of expected improvement acquisition function (Eq.~\ref{eq:ei}) and the exploration-exploitation trade-off. The surrogate model and the hidden target function are shown at top. The bottom shows the computed expected improvement $EI(x)$ of the surrogate model normalized for different choices of $\varepsilon$. A lower $\varepsilon$ favours exploitation near the current maximum while a higher $\varepsilon$ favours exploration and promotes sampling from areas with high uncertainty.}
\label{fig:ei}
\end{figure}

where $\mu$ and $\sigma$ are the mean and standard deviation of the probability distribution computed by the surrogate function, $\mu^+$ is the best observed point, $\phi(.)$ and $\varphi(.)$ are the probability density function (PDF) and cumulative distribution functions (CDF) of the Gaussian distribution, respectively. $\varepsilon$ determines the amount of exploration during the optimization with higher $\varepsilon$ leading to more exploration but less exploitation as shown in Figure~\ref{fig:ei}
for one dimension. EI is computed over the parameter space and the point with the maximum EI is sampled and is then used to update the objective function. A  detailed tutorial of Bayesian optimization can be found in~\cite{bo2}.

\paragraph{Implementation}\textit{ }\\
The COSY facility uses the Experimental Physics and Industrial Control System (EPICS)~\cite{epics}. This allows to fully automate the execution of the optimization method. The framework of executing the optimization method and controlling the IBL, has been implemented in Python and supporting libraries \cite{pyepics,cpymad,pybo,scikit-learn}. The supporting simulations have been performed in MAD-X~\cite{mad}. In order for the optimization method to work properly, a fast convergence of the beam and an accurate efficiency measurement are required. As the readings of the stripping foil and the beam current inside COSY are noisy, we increased the accuracy of the readings by averaging over several injection cycles. During the optimization process, there are many combinations of magnet currents that are not feasible, because they would lead to an immediate  loss of beam. As a reliable simulation model exists for the IBL, the combination of magnet currents that are quite distant from the acceptable values could be identified and omitted. This process will reduce the number of optimization steps and speeds up the convergence. We have adopted this approach during the live optimization of the IBL, albeit it is an optional step and should not change the final outcome. 

\begin{algorithm}[t]
\caption{Implementation of Bayesian optimization at COSY}\label{alg:bo}
$\mathcal{D} \gets \textit{empty list of tuples }(\textbf{x}, f(\textbf{x}))\\
r, \kappa \gets \textit{radius of the IBL beam pipe},\textit{factor threshold of beam width}\\
\mathcal{S} \gets \textit{empty list for the Twiss parameter }\beta_{x,y} \textit{ of all IBL components}\\
\textbf{Loop till convergence}: \\
\indnt \textbf{x}_t \gets \textit{currents of optimizable magnets} \\
\indnt f(\textbf{x}_t) \gets \textit{reading of the target current } \\
\indnt \mathcal{D} \gets \mathcal{D} \bigcup (\textbf{x}_t, f(\textbf{x}_t)) \\
\indnt \textbf{Loop}: \textit{ //search for the next valid parameters} \\ 
\indnt\indnt \textbf{x}_{t+1} \gets \textit{Bayesian optimization step over }\mathcal{D}\\
\indnt\indnt \textit{Simulation magnets} \gets \textbf{x}_{t+1}\\
\indnt\indnt \mathcal{S} \gets \textit{retrieve }\beta \textit{ values from the simulation Twiss parameters}\\
\indnt\indnt \beta_{\max} \gets \max\mathcal{S} \\
\indnt \textbf{Until }(\sqrt{\varepsilon\beta_{\max}} < \kappa\cdot r) 
\textit{ //verify the peak width is within the threshold} \\
\indnt \textit{Magnet currents at the IBL} \gets \textbf{x}_{t+1}\\
\textbf{Return }\arg\max_{\textbf{x}}\mathcal{D}
$
\end{algorithm}

A single optimization cycle starts with reading the currents of the magnets and the current of the target component. The magnet currents are the parameters to be optimized and the target component is either a Faraday cup, the strip foil current, or the beam current inside the storage ring. The parameters and the target value are added to the list buffer of observed parameters and targets. A single Bayesian optimization step is then executed by sequentially computing the surrogate function over the buffer and the acquisition function. The parameters of the maximum over the acquisition function represents the potential magnet currents to be explored. To reduce the number of optimization steps and roll-out parameters that are far off, the simulation model is updated with the computed parameters and the simulation is then executed. Using the Twiss parameters, the maximum beam size over the IBL is calculated and it is tested not to exceed the size of the IBL by a predefined threshold factor $\kappa$ which in our experiments was set to 4. If it exceeds the threshold width, the parameters of the beam are added to the buffer along with the target set to zero and a subsequent Bayesian optimization step is executed. If the parameters are in the allowed region, the IBL magnets are set to the computed values. Pseudocode for the full implementation algorithm is shown in Algorithm~\ref{alg:bo}. The space of acceptable parameters that leads to some beam passing through the IBL is very limited with respect to the full space of possible parameters. Therefore, the hyperparameters of the optimization are tuned towards more exploitation by setting $\varepsilon \approx 0$.

\section{Results}\label{sec:res}
Provided with sufficient time and fair parameters for the exploration-exploitation trade-off, Bayesian optimization typically continues to improve the optimization parameters. Experimentally, Bayesian optimization achieved better results on space dimension of size around 10 magnets or less compared to a much larger space. 
Therefore in the experiments described here we split the optimization
 into two consecutive parts. The first two segments, (I and II in figure~\ref{fig:cosy-ibl})
 with 11 quadrupole families are jointly optimized during the first step. The 24 steerers in these two sections were not part of the optimization.
 The third segment with four quadrupoles and four steerers was optimized during the second step. This resulted in a better convergence than optimizing all the 19 quadrupoles and steerers at once.

\begin{figure}
\centering
\includegraphics[width=.9\textwidth]{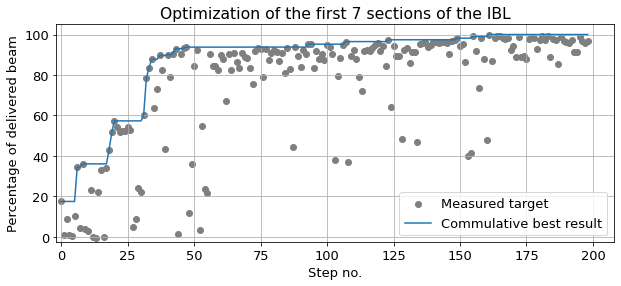}
\caption{Optimizing the quadrupole magnets of the first seven sections from the IBL via Bayesian optimization. The target is the reading of the Faraday beam cup between the last two sections and the parameters space is of size $d=11$. The best settings achieved full transmission of the beam from the start of the IBL till the target. This optimization run lasted for 50 minutes.}
\label{fig:bo-s12}
\end{figure}

\begin{figure}
\centering
\includegraphics[width=.9\textwidth]{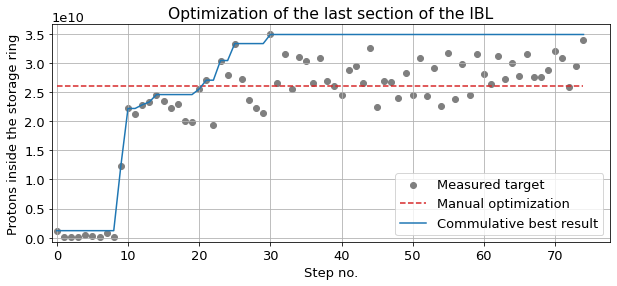}
\caption{Optimizing the quadrupole and steering magnets of the last section from the IBL via Bayesian optimization. After manual optimization of the IBL there were $2.6\times10^{10}$ protons inside the synchrotron. The objective function for Bayesian optimization is the beam current inside COSY and the parameters space is of size $d=8$. The beam intensity after the optimization improved to $3.5\times10^{10}$ protons inside the synchrotron and the duration of this optimization was 19 minutes.}
\label{fig:bo-s3}
\end{figure}

Figure~\ref{fig:bo-s12} shows a live run of Bayesian optimization for the first 
two segments (i.e. first seven sections) of the IBL, where full transmission of the beam up to the Faraday cup at the beginning of the last section
is achieved within 200 steps over 50 minutes.
Using a manual optimization for the last section of the IBL,
$2.6\times10^{10}$ protons inside the synchrotron were reached.
Bayesian optimization was executed subsequently to optimize the quadrupoles and steerers of the last section resulting in an increase of the intensity inside the synchrotron to $3.5\times10^{10}$ protons in 75 steps over 19 minutes as shown in Figure~\ref{fig:bo-s3}. 
Overall, the use of Bayesian optimization in this context demonstrated its effectiveness in improving the performance of the injection process. A manual optimization typically takes a few hours.

\section{Summary and Outlook}
This paper presents the optimization of the IBL at COSY using Bayesian optimization approach. This method has shown to facilitate and reduce the tuning time compared to manual optimization methods. In addition, the use of Bayesian optimization also showed the ability to increase the beam intensity compared to manual optimization. This demonstrates the potentials of machine learning algorithms in accelerator physics, particularly in the optimization of complex non-linear problems with multi-dimensional parameter spaces. Further research can be conducted on the use of other machine learning techniques that are more efficient during the online run but require extensive training such as Reinforcement Learning agent to enhance the performance and efficiency of accelerators.

\bibliography{bibtexBOcosy}
\bibliographystyle{IEEEtran}

\end{document}